\documentclass[amsmath,amssymb,aps,10pt,prd,twocolumn,letterpaper,nofootinbib,balancelastpage,notitlepage,superscriptaddress,floatfix,preprintnumbers]{revtex4-2}
\usepackage{graphicx}	
\usepackage{ragged2e}	
\usepackage{bm}		    
\usepackage{empheq}     

\newcommand{\be}{\begin{equation}}
\newcommand{\ee}{\end{equation}}

\newcommand{\mAp}{m_{A^\prime}}

\newcommand{\eps}{\epsilon}

\newcommand{\p}{\prime}

\newcommand{\lap}{\nabla}

\newcommand{\order}[1]{\mathcal{O}{(#1)}}
\newcommand{\w}{\omega}

\newcommand{\g}{\gamma}

\newcommand{\E}{\boldsymbol{E}}

\newcommand{\xv}{{\bf x}}

\usepackage[sort&compress]{natbib}	 	
	\setcitestyle{square,numbers,comma}	
\usepackage[colorlinks=true,urlcolor=blue,linkcolor=red,citecolor=red]{hyperref}	

\usepackage{mwe}
\usepackage{graphicx}
\usepackage{subcaption}
\usepackage{caption}
\usepackage{soul}

\newcommand{\tabref}[2][]{Table{#1}~\ref{tab:#2}}		
\newcommand{\figref}[2][]{Fig{#1}.~\ref{fig:#2}}		
\renewcommand{\eqref}[2][]{Eq{#1}.~(\ref{eq:#2})}		
\newcommand{\eqrefRange}[2]{Eqs.~(\ref{eq:#1})--(\ref{eq:#2})}		
\newcommand{\citeR}[2][]{Ref{#1}.~\cite{#2}}			
\newcommand{\Hz}[1][]{\,\mathrm{{#1}Hz}}
\newcommand{\s}[1][]{\,\mathrm{{#1}s}}

\usepackage{ezedits}
\defineEdit{SK}{\color[rgb]{0.01, 0.75, 0.24}}{\color[rgb]{0.01, 0.75, 0.24}}
\defineEditQuick{HC}{blue}
\defineEditQuick{ZL}{magenta}
\defineEditQuick{AB}{cyan}
\defineEditQuick{AH}{purple}
\defineEditQuick{RH}{blue}
\defineEditQuick{BG}{blue}
\defineEditQuick{AM}{red}

\begin{document}

\title{Improved Dark Photon Sensitivity from the Dark SRF Experiment}
\date{\today}
\author{Saarik Kalia}
\email{kalias@umn.edu}
\affiliation{School of Physics \& Astronomy, University of Minnesota, Minneapolis, MN 55455, USA}
\author{Zhen Liu}
\email{zliuphys@umn.edu}
\affiliation{School of Physics \& Astronomy, University of Minnesota, Minneapolis, MN 55455, USA}
\author{Bianca Giaccone}
\affiliation{Superconducting Quantum Materials and Systems Center (SQMS) Division, Fermi National Accelerator Laboratory, Batavia, 60510, IL, USA}
\author{Oleksandr Melnychuk}
\affiliation{Superconducting Quantum Materials and Systems Center (SQMS) Division, Fermi National Accelerator Laboratory, Batavia, 60510, IL, USA}
\author{Roman Pilipenko}
\affiliation{Superconducting Quantum Materials and Systems Center (SQMS) Division, Fermi National Accelerator Laboratory, Batavia, 60510, IL, USA}
\author{Asher Berlin}
\affiliation{Theoretical Physics Division, Fermi National Accelerator Laboratory, Batavia, 60510, IL, USA}
\author{Anson Hook}
\affiliation{Maryland Center for Fundamental Physics, Department of Physics, University of Maryland, College
Park, MD 20742, U.S.A.}
\author{Sergey Belomestnykh}
\affiliation{Applied Physics and Superconducting Technology Directorate, Fermi National Accelerator Laboratory, Batavia, 60510, IL, USA}
\author{Crispin Contreras-Martinez}
\affiliation{Applied Physics and Superconducting Technology Directorate, Fermi National Accelerator Laboratory, Batavia, 60510, IL, USA}
\author{Daniil Frolov}
\affiliation{Superconducting Quantum Materials and Systems Center (SQMS) Division, Fermi National Accelerator Laboratory, Batavia, 60510, IL, USA}
\affiliation{Present address: IBM Research, Yorktown Heights, 10598, NY, USA}
\author{Timergali Khabiboulline}
\affiliation{Applied Physics and Superconducting Technology Directorate, Fermi National Accelerator Laboratory, Batavia, 60510, IL, USA}
\author{Yuriy Pischalnikov}
\affiliation{Applied Physics and Superconducting Technology Directorate, Fermi National Accelerator Laboratory, Batavia, 60510, IL, USA}
\author{Sam Posen}
\affiliation{Applied Physics and Superconducting Technology Directorate, Fermi National Accelerator Laboratory, Batavia, 60510, IL, USA}
\author{Oleg Pronitchev}
\affiliation{Superconducting Quantum Materials and Systems Center (SQMS) Division, Fermi National Accelerator Laboratory, Batavia, 60510, IL, USA}
\author{Vyacheslav Yakovlev}
\affiliation{Superconducting Quantum Materials and Systems Center (SQMS) Division, Fermi National Accelerator Laboratory, Batavia, 60510, IL, USA}
\author{Anna Grassellino}
\email{annag@fnal.gov}
\affiliation{Superconducting Quantum Materials and Systems Center (SQMS) Division, Fermi National Accelerator Laboratory, Batavia, 60510, IL, USA}
\author{Roni Harnik}
\email{roni@fnal.gov}
\affiliation{Theoretical Physics Division, Fermi National Accelerator Laboratory, Batavia, 60510, IL, USA}
\author{Alexander Romanenko}
\email{aroman@fnal.gov}
\affiliation{Applied Physics and Superconducting Technology Directorate, Fermi National Accelerator Laboratory, Batavia, 60510, IL, USA}
\preprint{UMN-TH-4508/25}
\preprint{FERMILAB-PUB-25-0657-SQMS-TD}

\begin{abstract}
We report the refined dark-photon exclusion bound from Dark SRF's pathfinder run. Our new result is driven by improved theoretical modeling of frequency instability in high-quality resonant experiments. Our analysis leads to a constraint that is an order of magnitude stronger than previously reported (corresponding to a signal-to-noise ratio that is four orders of magnitude larger).  This result represents the world-leading constraint on non-dark-matter dark photons over a wide range of masses below $6\,\rm \mu eV$ and translates to the best laboratory-based limit on the photon mass $m_\gamma<2.9\times 10^{-48}\,\rm g$.  
\end{abstract}

\maketitle

{\flushleft\textbf{Introduction.--}}
A number of outstanding problems in fundamental physics indicate the need for new particles beyond the Standard Model (SM).  One of the simplest extensions which can be made to the SM is the addition of a new $U(1)$ gauge symmetry, mediated by a new vector particle, often dubbed a ``dark photon".  Such additional symmetries are predicted by a number of theoretical models~\cite{cvetivc1996implications,Nelson:2011sf,Graham:2015rva}.  The dark photon may be able to convert to/from the SM photon through a small kinetic mixing interaction~\cite{Holdom:1986ag}, and several experiments have been designed to search for it~\cite{Redondo_2008,ehret2010new,Betz_2013,Parker_2013,Kroff_2020,handbook}.

One of the most sensitive experiments searching for the existence of dark photons is the Dark SRF experiment~\cite{Romanenko2023}, which utilizes high-$Q$ superconducting radio frequency (SRF) cavities to employ a ``light-shining-through-walls" (LSW) search for dark photons~\cite{Okun:1982xi,Bibber1987,JAECKEL2008509,Graham:2014sha}.  The Dark SRF setup consists of an emitter and a receiver cavity, both tuned to the same resonant frequency $f_0\sim1.3\Hz[G]$, with quality factors of $Q\sim10^{10}$.  SM photons are injected into the emitter cavity.  In the presence of a kinetic mixing interaction, these SM photons can convert to dark photons, which may reconvert to SM photons inside the receiver cavity.  A measurement of anomalous power in the receiver cavity would therefore indicate the existence of a dark photon.  Notably, this experiment does not rely on the assumption that the dark photon constitutes dark matter (DM).

\citeR{Romanenko2023} reported the first results from the Dark SRF pathfinder run.  In the absence of a detection, they set world-leading constraints on non-DM dark photons and a competitive constraint on the photon mass.  The Dark SRF cavities exhibited temporal variations in their resonant frequencies.  These included both a slow secular drift and fast stochastic jittering of the resonant frequency.  The latter effect, known as ``microphonics" or jittering, can arise from nanometer-scale deformations of the cavity walls, e.g. due to bubble collisions from the cooling fluid.  \citeR{Romanenko2023} took a conservative approach in modeling these effects, by assuming that the emitter and receiver frequencies were always mismatched by a frequency difference of order the amplitude of the microphonics.  This resulted in a suppression of $\sim10^{-5}$ in the expected signal power in the receiver cavity.  In an accompanying regular article~\cite{cui2025}, we analyzed the spectral response of a jittering resonator with a more accurate modeling of frequency jittering, and showed that jittering does not always suppress power accumulation in resonant systems.  In particular, we found that power suppression is small in the case of Dark SRF.  With the new understanding from~\citeR{cui2025}, in this work, we develop a new analysis of the Dark SRF's pathfinder run to show that its existing data translates to much more powerful constraints on a dark-photon kinetic mixing and a photon mass than originally reported in \citeR{Romanenko2023}.

\begin{figure*}[htbp]
    \centering
    \includegraphics[width=0.7
    \textwidth]{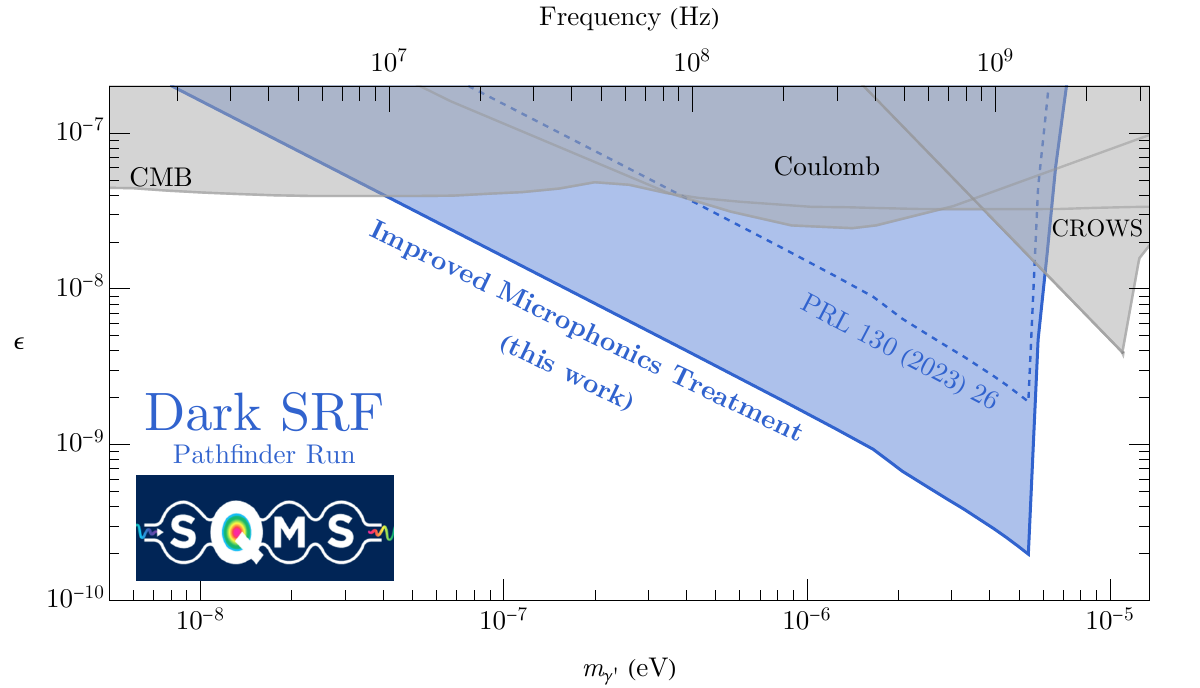}
    \caption{Refined dark-photon exclusion bound, based on the Dark SRF pathfinder run, utilizing our improved jittering treatment. Our new result (solid line) improves the published result (dashed line)~\cite{Romanenko2023} by about an order of magnitude. See the text for details.
    }
    \label{fig:DarkSRF}
\end{figure*}

This work is organized as follows. We first summarize the relevant main findings of \citeR{cui2025}, including the conditions under which jittering leads to power suppression.  We then show our re-analysis of the Dark SRF pathfinder run, in light of this improved modeling of jittering.  The refined Dark SRF bound, shown in \figref{DarkSRF}, is an order of magnitude stronger than previously reported. We further demonstrate how this bound translates to a world-leading laboratory-based limit on the photon mass.  Finally, we conclude with a brief discussion on upcoming improvements to the Dark SRF experiment.

{\flushleft\textbf{Jittering resonator.--}}
Understanding the power accumulation in a jittering resonator and the resulting sensitivity of the system are essential for high-Q sensing experiments.  Importantly, the accompanying regular article~\cite{cui2025} showed that if the jittering is sufficiently fast,
then the resulting power suppression is minimal.  In the case that the jittering is exactly monochromatic, the frequency of the jittering must be larger than the amplitude of the jittering.  More generally, the condition is given by \eqrefRange{alpha}{rho}.  In particular, this condition is satisfied for the Dark SRF experiment, so that jittering should not significantly impact the exclusion bound that Dark SRF sets on dark-photon parameter space.

{\flushleft\textit{Modeling.--}}
The accompanying article~\cite{cui2025} modeled a general resonant system whose natural frequency exhibits stochastic variations over time.  Generically, such a system is described by the stochastic differential equation
\begin{equation}
    \ddot x(t) + \gamma \dot x(t) + (\omega_0 +\delta \omega (t))^2 x(t) = F(t),
    \label{eq:resonator}
\end{equation}
where $x(t)$ denotes the amplitude of the resonator, $\gamma$ its linewidth, $\omega_0$ its central natural frequency, $\delta\omega(t)$ its stochastic variations, and $F(t)$ a driving force on the system.  In the case of Dark SRF, $x(t)$ represents the electric field amplitude in the cavity, while $F(t)$ represents external currents which can drive the system.  These could include either real currents which constitute thermal noise, or an effective current resulting from a dark photon.  In the latter case, the force would be (nearly) monochromatic, while in the former it would be broadband.  We will consider both of these cases in the subsequent subsections.

The jittering $\delta\omega(t)$ is modeled as a stochastic process described by a power spectrum $S_{\delta\omega}(\omega)$.  The empirical jittering spectrum of the SRF cavities used in the Dark SRF experiment was measured in \citeR{pischalnikov2019operation}.  The spectrum exhibits a dominant peak, so for simplicity, we consider a Lorentzian spectrum for the jittering
\begin{equation}
    S_{\delta\omega}(\omega)=\frac{\delta\omega_0^2}{\tau}\left(\frac1{\tau^{-2}+(\omega-\omega_j)^2}+\frac1{\tau^{-2}+(\omega+\omega_j)^2}\right),
    \label{eq:Sdelomega}
\end{equation}
where $\delta\omega_0$ represents the typical amplitude of jittering, $\omega_j$ its peak frequency, and $2/\tau$ the linewidth of the peak.  \tabref{parameters} shows the values of these parameters, as well as $\omega_0$ and $\gamma$, used to model the Dark SRF cavities.

\begin{table}
    \centering
    \begin{tabular}{c c c}
        \hline\hline
        Parameter&Symbol&Value\\
        \hline
        Central natural frequency&$f_0=\frac{\omega_0}{2\pi}$\textsuperscript{\ref{ftnt:labels}}&$1.3\,\mathrm{GHz}$\\
        Resonator linewidth&$\gamma$&$2\pi\times0.15\Hz$\\
        Jittering amplitude&$\delta f_0=\frac{\delta\omega_0}{2\pi}$&$3\Hz$\\
        Jittering correlation time&$\tau$&$0.064\s$\\
        Peak jittering frequency&$f_j=\frac{\omega_j}{2\pi}$&$45\Hz$\\
        \hline\hline
    \end{tabular}
    \caption{Benchmark parameters for the Dark SRF experiment, based on \citeR[s]{Romanenko2023,pischalnikov2019operation}.  The receiver cavity is modeled as a jittering resonator, which satisfies \eqref{resonator}.  The jittering is modeled as a random process whose PSD has a single Lorentzian peak, as in \eqref{Sdelomega}.  The values of $f_j$ and $\tau$ above correspond to the main peak in the jittering spectrum measured by \citeR{pischalnikov2019operation}, which exhibited a central frequency of $45\Hz$ and linewidth of $5\Hz$.
    }
    \label{tab:parameters}
\end{table}

\footnotetext{\label{ftnt:labels}%
Throughout this work, we will refer to both the ordinary frequency $f_\mathrm{label}$ and angular frequency $\omega_\mathrm{label}$ for various quantities.  These are always related by $\omega_\mathrm{label}=2\pi f_\mathrm{label}$.}

The power spectrum $S_{\delta\omega}$ specifies the two-point statistics of the jittering.  In \citeR{cui2025}, we considered multiple models of the higher point statistics.  Importantly, we showed that when the power suppression from jittering is small (as is the case for Dark SRF), the results are insensitive to these higher point statistics.

{\flushleft\textit{Power accumulation.--}}
Let us first review the effect of jittering on power accumulation.%
\footnote{In this subsection, we refer to $|x|^2$ as the ``power" in the resonator.  If $x$ represents the electric field amplitude of a cavity mode, then $|x|^2$ is proportional (although not equal) to the energy in the mode.  In this sense, $|x|^2$ is a proxy for the time-averaged power that would be extracted by a readout.  (We complexify $x$, so that $|x|^2$ is not oscillatory.)}
In this subsection, we consider an on-resonance monochromatic force $F(t)=F_0e^{i\omega_0t}$.  In the absence of jittering, $\delta\omega(t)=0$, a resonator driven by such a force will first experience transient behavior based on its initial conditions.  The asymptotic behavior of the system will, however, be independent of the inital conditions.  Specifically, after the transient behavior has decayed away, the power in the resonator will be given by 
\begin{equation}
    |x_0(\infty)|^2\equiv\left.\lim_{t\rightarrow\infty}|x(t)|^2\right|_{\delta\omega(t)=0}=\frac{|F_0|^2}{\gamma^2\omega_0^2}.
    \label{eq:nojittering_power}
\end{equation}
If  there is a fixed frequency offset between the driving force and resonant frequency, $\delta\omega(t)=\delta\omega_0$, the power will instead be given by
\begin{align}
    |x_\mathrm{fix}(\infty)|^2&\equiv\left.\lim_{t\rightarrow\infty}|x(t)|^2\right|_{\delta\omega(t)=\delta\omega_0}\\
    &=\frac{\gamma^2}{\gamma^2+4\delta\omega_0^2}\cdot\frac{|F_0|^2}{\gamma^2\omega_0^2}.
    \label{eq:suppression}
\end{align}
This prefactor represents the maximum power suppression that jittering of typical amplitude $\delta\omega_0$ can result in.  This was the power suppression used to compute the original dark-photon exclusion bound presented in \citeR{Romanenko2023}.  For the Dark SRF experiment, this prefactor is $\sim10^{-5}$.

Detailed in the accompanying regular article~\cite{cui2025}, we performed a perturbative calculation of the expected asymptotic power in the presence of nonzero jittering.  Specifically, our calculation applied when $\alpha\ll1$, where
\begin{align}\label{eq:alpha}
    \alpha&\equiv\frac{4\delta\omega_0^2}{\gamma^2}\rho,\\
    \rho&\equiv\frac{\gamma\tau(2+\gamma\tau)}{(2+\gamma\tau)^2+4\omega_j^2\tau^2}.
\label{eq:rho}\end{align}
In this case, the expected power is given by
\begin{equation}
    \langle|x(t)|^2\rangle_\infty\approx\frac{|x_0(\infty)|^2}{1+\alpha},
    \label{eq:power}
\end{equation}
where $\langle\cdot\rangle_\infty$ represents both the $t\rightarrow\infty$ limit and an ensemble average over realizations of the jittering.

The quantity $\rho$ highlights the dependence of the results on the relevant timescales of the jittering.  When $\tau\rightarrow\infty$ and $\omega_j\rightarrow0$, then $\rho\rightarrow1$ and \eqref{power} recovers the worst-case suppression in \eqref{suppression}.  This implies that when the jittering is slow, the power accumulation is inhibited as if the system were always off-resonance.  However, when $\tau\rightarrow0$ or $\omega_j\rightarrow\infty$, then $\rho\rightarrow0$ and \eqref{power} recovers the unsuppressed power in \eqref{nojittering_power}.  This implies that when the jittering is fast, the system accumulates power as if there were no jittering at all!  In the case of the Dark SRF experiment, we evaluate $\alpha\approx0.15$, so that \eqref{power} predicts the experiment should accumulate roughly 87\% of the power it would have in the absence of jittering.

This result can be intuitively understood in terms of the relative phase $\theta(t)$ between the resonator and driving force.  It is not difficult to show that \eqref{resonator} [with a monochromatic driving force] implies
\begin{align}
    \label{eq:accumulation}
    \frac d{dt}|x(t)|^2&=-\gamma|x(t)|^2+\frac{|F_0||x(t)|}{\omega_0}\sin\theta(t),\\
    \theta(t)&=\arg\left[x(t)^*F_0e^{i\omega_0t}\right].
\end{align}
The first term in \eqref{accumulation} represents power lost through dissipation, while the second term represents power gained/lost due to the driving force, which depends on its relative phase $\theta(t)$ with the resonator.  In the absence of jittering, a resonant driving force leads to asymptotic behavior with $\theta=\pi/2$, so that the system gains maximal power from the driving force.  An off-resonance force leads to deviations from $\theta=\pi/2$, meaning the system does not gain power efficiently and so the asymptotic power is suppressed.  Likewise, jittering also allows the relative phase to evolve and deviate from its optimal value.  If the jittering is slow, a large relative phase can accumulate.  However, if the jittering is fast, the relative phase will be washed out by the rapidly changing sign of the jittering.  As a result, power accumulation will proceed as in the on-resonance case.

{\flushleft\textit{Sensitivity.--}}
The results above demonstrate that jittering does not significantly affect the power that the Dark SRF receiver cavity would accumulate in the presence of a signal.  We would like to understand how this affects the overall sensitivity of the experiment.  In \citeR{Romanenko2023}, the sensitivity of the Dark SRF experiment was characterized by a signal-to-noise ratio (SNR), given by
\begin{equation}
    \mathrm{SNR}=\frac{P_\mathrm{rec}}{P_\mathrm{th}}\sqrt{\delta\nu\,t_\mathrm{int}}.
    \label{eq:snr}
\end{equation}
Here $P_\mathrm{rec}$ and $P_\mathrm{th}$ represent the power that the system would accumulate from a signal or from thermal noise, respectively.  Meanwhile, $t_\mathrm{int}$ is the total integration time of the experiment, and $\delta\nu$ represents the bandwidth of the response.  The results described in the previous section imply that jittering only reduces $P_\mathrm{rec}$ by a factor of $(1+\alpha)^{-1}\approx0.87$. (In the Dark SRF experiment, $P_\mathrm{th}$ is dominated by extrinsic noise, and so is not degraded by this factor.)

The bandwidth $\delta\nu$ corresponds to the frequency range over which the SNR of the system is maximized.  This is typically the smaller of two quantities: the linewidth of the signal (in this case, the dark-photon effective current sourced by the emitter cavity), or the frequency range over which thermal noise dominates over readout noise.
The jittering of the receiver cavity only affects the latter quantity.  \citeR{cui2025} also analyzes the spectral response of a jittering resonator, and shows that when $\alpha\ll1$, the lineshape of the response near the resonant peak is not modified.  Therefore, we deduce that $\delta\nu$ is unaffected by the jittering of the receiver cavity.%

The central frequency of the emitter cavity also experiences jittering.  The full Dark SRF experiment, therefore, consists of two jittering resonators: the emitter cavity $x_\mathrm{emit}(t)$, which is driven by an external power source $F_\mathrm{emit}(t)$; and the receiver cavity $x_\mathrm{rec}(t)$, which is driven by the dark-photon effective current $F_\mathrm{rec}(t)$.  As mentioned above, the linewidth of $F_\mathrm{rec}(t)$ can affect $\delta\nu$, and so we must consider the impact of the emitter's jittering on this linewidth.  The jittering of the emitter cavity directly affects how $x_\mathrm{emit}(t)$ accumulates power from $F_\mathrm{emit}(t)$.  The above-mentioned spectral results indicate that the emitter jittering does not influence the linewidth of $x_\mathrm{emit}(t)$ [in the perturbative regime $\alpha\ll1$].  As $F_\mathrm{rec}(t)$ ultimately inherits its spectrum from $x_\mathrm{emit}(t)$, we conclude that the jittering of the emitter cavity can be ignored.

In conclusion, we find that \eqref{snr} can be readily applied to compute the sensitivity of Dark SRF, with the only modification from jittering being a slight reduction in $P_\mathrm{rec}$.

{\flushleft\textbf{Improved Dark SRF Pathfinder Result.--}}
In \citeR{Romanenko2023}, for the first result of the Dark SRF pathfinder run, a dark-photon exclusion bound was calculated, assuming a power suppression factor of $7.7\times10^{-6}$ due to frequency drifting and jittering.  In the previous section, we showed that the true suppression is only $0.87$!  The existing Dark SRF data, therefore, translates to a much stronger constraint than previously reported.  In \figref{DarkSRF}, we show the refined dark-photon exclusion bound from Dark SRF, which properly accounts for jittering.  To derive this refined bound, we utilize the measured spectrum from the full 35-minute Pathfinder run. To avoid effects from the slow secular drift of the resonant frequency, we conservatively assume that the dark-photon signal only appears in the first 150\,s of the data. A period of 150\,s was chosen to ensure that the central frequency did not drift more than $\gamma/(2\pi)$, based on the reported secular drift rate of $5.7\Hz$ over 100\,min~\cite{Romanenko2023}.  (The emitter and receiver frequencies were set to match at the beginning of the run.)  We also exclude the first 15\,s of power accumulation to remove transient contributions.  
Therefore, the current analysis makes the following modifications to the original Dark SRF analysis: replacing the suppression factor of $7.7\times10^{-6}$ with the appropriate $0.87$ factor derived from our jittering analysis; and rescaling the dark-photon signal by $150\,\mathrm{s}/35\,\mathrm{min}$ to account for the reduction in signal integration time (without correspondingly rescaling the thermal noise).
Remarkably, this refined jittering treatment already improves the sensitivity to the kinetic mixing parameter $\epsilon$ by about an order of magnitude. 
Note that the sensitivity to $\epsilon$ scales as $\sim\mathrm{SNR}^{1/4}$ in LSW searches, and so this translates to an enhancement of the SNR by four orders of magnitude!

In \figref{DarkSRF}, we also show existing constraints on non-DM dark photons from: the CROWS cavity-based LSW experiment~\cite{Betz_2013}, tests of Coulomb's law~\cite{Williams:1971ms,Abel_2008}, and CMB spectral distortions~\cite{McDermott_2020} (see also \citeR[s]{Arsenadze_2025,chluba2024}).
This improved Dark SRF pathfinder-run result is the world-leading constraint on non-DM dark photons over a wide range of masses below $6\,\rm \mu eV$.
In future runs, Dark SRF will mitigate the drifting of the receiver cavity frequency, so that a longer integration time can be used, and a higher sensitivity can be achieved.
This improved jittering treatment will enable Dark SRF to fully utilize the potential of high-$Q$ devices for new-physics searches.

{\flushleft\textbf{World-Leading Laboratory-based limits on Photon Mass.--}}
The possibility that the SM photon has a small mass, or equivalently that Coulomb's law deviates from the observed inverse square force law,  is one of the most well-motivated extensions to particle physics. Indeed, aside from electromagnetism and gravity, every observed force has been found to deviate from the inverse square law.

The most powerful laboratory bounds on the photon mass arise from tests of Coulomb's inverse square law (see, e.g., Refs.~\cite{Tu:2005ge,GSpavieri_2004} for a review). A well known example is the so-called Cavendish test, first performed in the mid-18th century by Cavendish and a hundred years later by Maxwell~\cite{maxwell}. In such an experiment, an electric field is measured inside of a charged shell, whose presence would signal the violation of Coulomb's and Gauss's laws. Since then, more recent studies have implemented improved setups~\cite{Plimpton:1936ont,Bartlett:1970js,Cochran1968,Williams:1971ms,Crandall:1983ec}, with the most sensitive of these bounding the photon mass to be lighter than $m_\g \lesssim 6.4 \times 10^{-15} \ \text{eV} = 1.1 \times 10^{-47} \ \text{g}$~\cite{Williams:1971ms}.

The signal in such an experiment can be most simply described by noting that  Gauss's and Amp\`{e}re's laws are modified to include additional source terms, described by the effective four-current $j_\text{eff}^\mu \equiv - m_\g^2 \, A^\mu$. Note that this is similar to the effective current arising from a kinetically-mixed dark photon $j_\text{eff}^\mu = - \eps \, \mAp^2 \, A^{\p \mu}$~\cite{Graham:2014sha,Chaudhuri:2014dla}. Since the dark photon field emitted by a visible source is $A^{\p \mu} \sim \order{\eps} \, A^\mu$, this suggests the simple mapping
\be
\label{eq:mapping}
m_\g \leftrightarrow \eps \, \mAp
\ee
can be used to recast the sensitivity of a kinetically-mixed dark photon to the mass of the SM photon. Before we explicitly verify \eqref{mapping} below, we note that taking it at face value implies that the dark photon Dark SRF 95\% C.L. limit can be recast as
\be
\label{eq:DarkSRFProca}
m_\g \lesssim 1.6 \times 10^{-15} \ \text{eV} = 2.9 \times 10^{-48} \ \text{g}
~.
\ee
This overcomes the previous best laboratory limit on $m_\g$ (as obtained by the Cavendish test in Ref.~\cite{Williams:1971ms}) by a factor of $\sim 4$. Although stronger limits have been derived from the consideration of magnetic fields on solar and galactic length scales~\cite{Tu:2005ge}, these are subject to systematic uncertainties in the modeling of astrophysical systems and can even be evaded completely in Higgsed models~\cite{Adelberger:2003qx}. 

Now, let us provide a more careful derivation of \eqref{mapping}, which shows that this holds for searches that are dominantly sensitive to the longitudinal mode of the dark photon, such as Dark SRF. 
The intuition for this statement is that to leading order in the photon mass, $m_\gamma/\omega$, the longitudinal mode of the photon does not couple to charged particles and acts exactly like the longitudinal mode of a dark photon.  In this sense, a longitudinal mode is a longitudinal mode; it does not matter if it arises from the mass of a kinetically-mixed dark photon or the mass of our very own photon. 

Since the longitudinal mode of the photon decouples when the mass goes to zero, its couplings must vanish in the $m_\gamma \rightarrow 0 $ limit. Thus, we expect it to penetrate through an experimental setup analogous to a dark photon field. This can be seen from the wave equation for a longitudinal wave, supplemented with Ohm's Law, which implies that the longitudinal skin-depth in a material of conductivity $\sigma \gg \w \gg m_\g$ is $\delta \sim \sigma / m_\g^2$. 

Let us now consider the production of the longitudinal mode of a massive photon.  In doing so, we perform a Taylor series of the electric field in powers of $m_\g^2$, i.e., $\E = \sum_{n=0}^\infty \E^{(n)}$, where $\E^{(n)} \sim \order{m_\g^{2n}}$. At $\order{m_\g^0}$, the fields of the emitter cavity are unmodified,
$\E^{(0)} = \E_\text{em}^{(0)} \, e^{i \w t}$, such that the in-vacuum Proca wave equation $(\lap^2 + \w^2 - m_\g^2 ) \, \E = 0$ can be rewritten to $\order{m_\g^2}$ as $(\lap^2 + \w^2) \, \E^{(1)} = m_\g^2 \, \E^{(0)}$. Solving this using the Green's function, and keeping only the unscreened longitudinal component, we have
\be
\label{eq:EmProca}
\E^{(1)}_L (\xv) = - \, m_\g^2 \, \int d^3 \xv^\p ~ \frac{\E_\text{em}^{(0)} (\xv^\p)}{4 \pi \, | \xv - \xv^\p |} ~ e^{- i \w | \xv - \xv^\p |} \bigg|_L
~,
\ee
where the integral is performed over the volume of the emitter cavity and the subscript ``$L$" refers to a projection onto the longitudinal polarization. This thus contains the leading order form for the longitudinal photon mode as sourced by the emitter cavity, which is not screened by the conducting walls.

We can now discuss the response of the receiver cavity to the unscreened longitudinal mode sourced by the emitter cavity. To leading order in the photon mass, the response of the transverse electric field $\E_\text{rec}^{(1)}$ in the receiver cavity is described by~\cite{Graham:2014sha}
\be
\label{eq:RecProca}
\lap \times \lap \times \E_\text{rec}^{(1)} - \w^2 \E_\text{rec}^{(1)}  =  \lap \, \lap \cdot \E_L^{(1)}
~.
\ee
Above, the right-hand side involving the longitudinal mode appears as an effective source term for the receiver cavity. 

Comparing to the case of a kinetically-mixed dark photon,  we note that \eqref{EmProca} and \eqref{RecProca} are the same as the corresponding dark photon equations to leading order in the small coupling $\eps \, \mAp$ (see, e.g., Eq.~(24) and Eq.~(48) of Ref.~\cite{Graham:2014sha}), as long as one makes the identification in \eqref{mapping}.  This shows that the Dark SRF bound on the longitudinal mode of the dark photon can easily be recast as a bound on the mass of the SM photon, as in \eqref{DarkSRFProca}.

{\flushleft\textbf{Summary and Outlook.--}}
In this work, we undertook a refined analysis of the Dark SRF pathfinder run.  We incorporated a proper treatment of jittering effects in high-$Q$ cavities, adapting the main results detailed in \citeR{cui2025}. These considerations allowed us to derive a modified search strategy and improved dark-photon exclusion limit from the Dark SRF pathfinder run.  The original Dark SRF exclusion limit presented in \citeR{Romanenko2023} took a conservative approach in its modeling of frequency instability, and assumed that frequency drift and microphonics together resulted in five orders of magnitude of power suppression.  \citeR{cui2025} showed that, in fact, the parameters of Dark SRF lie in a perturbative regime where the impact of microphonics on power accumulation is small.  This allowed us to refine the resulting bound by an order of magnitude, as shown in \figref{DarkSRF}, by analyzing a subset of the data in which frequency drift was negligible.  These results translate to a world-leading limit on the photon mass, $m_\gamma<2.9\times10^{-48}\,\mathrm g$, among laboratory experiments.

The next-generation Dark SRF is being engineered for operation in a dilution refrigerator (DR) and will benefit from several improvements, in addition to the refined microphonics analysis. For compatibility with the DR, the cavity volume has been reduced and the frequency increased to $2.6\ \text{GHz}$. Additional improvements include a compact piezo-only tuning system~\cite{contreras-martinez:srf2023-wepwb133_tuner} for the emitter and a rigidly mounted receiver cavity. A proportional–integral control loop~\cite{contreras-martinez:srf2023-wepwb109_PILoop} stabilizes the emitter frequency to $0.1\ \text{Hz/hr}$ in liquid helium, with the receiver drifting at $1.7\ \text{Hz/hr}$. Microphonics in liquid helium show RMS levels of $2.6\ \text{Hz}$ (emitter) and $2.1\ \text{Hz}$ (receiver). The analysis in Ref.~\cite{contreras-martinez:srf2023-wepwb109_PILoop} indicates an optimal run integration time of $\approx 200\ \text{s}$ for maintaining peak sensitivity. Crosstalk mitigation now permits $15\ \text{MV/m}$ excitation with leakage below the $2\ \text{K}$ thermal background.
These upgrades, validated in liquid helium for rapid iteration, will require adaptation to the DR’s vibrational environment, where drift, microphonics, and emitter’s power dissipation will require characterization in situ. In the $\text{mK}$ regime, incorporating a Josephson parametric amplifier would further improve the SNR and the experiment’s ultimate sensitivity.

\acknowledgments
This work was supported by the U.S. Department of Energy, Office of Science, National Quantum Information Science Research Centers, Superconducting Quantum Materials and Systems Center (SQMS), under Contract No. 89243024CSC000002. Fermilab is operated by Fermi Forward Discovery Group, LLC under Contract No. 89243024CSC000002 with the U.S. Department of Energy, Office of Science, Office of High Energy Physics. S.K. and Z.L. are supported in part by the DOE grant DE-SC0011842 and a Sloan Research Fellowship from the Alfred P. Sloan Foundation at the University of Minnesota.  A.H. is supported by NSF grant PHY-2210361 and the Maryland Center for Fundamental Physics. The data for the final sensitivity curve in the figure is made publicly available through Github~\cite{github} under CC-BY-NC-SA.


\bibliographystyle{JHEP}
\bibliography{references.bib}

\end{document}